\documentclass[twocolumn,showpacs,amsmath,amssymb,prl]{revtex4}

\usepackage{graphicx,bm,bbm,amsfonts}

\newcommand{\bra}[1]{\langle #1|}
\newcommand{\ket}[1]{|#1\rangle}

\newcommand{\tr}[1]{{\,\rm tr}{\lbrack #1 \rbrack}\,}
\newcommand{\trr}[2]{{\,\rm tr_{#1}}{\lbrack #2 \rbrack}\,}
\def\ii{{\rm i}}

\def\etal#1{#1}
\def\etall#1{ {\em et.al.}}
\def\tit#1{}

\begin{document}

\title{Exact convergence times for generation of random bipartite entanglement}

\author{Marko \v Znidari\v c}

\affiliation{Department of Physics, Faculty of Mathematics and Physics, University of Ljubljana, SI-1000 Ljubljana, Slovenia}

\begin{abstract}
We calculate exact convergence times to reach random bipartite entanglement for various random protocols. The eigenproblem of a Markovian chain governing the process is mapped to a spin chain, thereby obtaining exact expression for the gap of the Markov chain for any number of qubits. For protocols coupling nearest neighbor qubits and CNOT gate the mapping goes to XYZ model while for U(4) gate it goes to an integrable XY model. For coupling between a random pair of qubits the mapping is to an integrable Lipkin-Meshkov-Glick model. In all cases the gap scales inversely with the number of qubits, thereby improving on a recent bound in~[Phys.~Rev.~Lett. {\bf 98}, 130502 (2007)].    
\end{abstract}
\maketitle

{\em Introduction--} Entanglement is a resource that can make quantum processes more powerful than classical ones. While a complete characterization of entanglement is a complicated task one might learn a great deal from properties of generic states, where by generic we mean states drawn randomly according to unitarily invariant Haar measure, shortly random states. Random states have almost maximal bipartite entanglement~\cite{max} and are a needed resource in some quantum information protocols like quantum dense coding~\cite{Harrow:04} or remote state preparation~\cite{rsp}. In addition, they occur naturally during a time evolution of a sufficiently complex (quantum chaotic) system. Note though that detecting entanglement in random states could be difficult~\cite{ranW}. Central issue in quantum information processes is their efficiency. We want protocols that are faster than the best classical. Natural question therefore is, how difficult is it to generate random quantum states? Knowledge about chaotic quantum systems might suggest that it should be possible to produce random states, as far as their bipartite entanglement is concerned, in polynomial time, {\em i.e.}, number of two qubit gates. On the other hand, one should be aware that to produce an arbitrary unitary transformation, and therefore a truly random state, an exponential number of two qubit gates is needed in general~\cite{Emerson:03}. However, if our criterion is just to reproduce bipartite entanglement of typical random states, which is the case frequently studied, less gates might be needed.  

Random protocol for generating random states is as follows: generate some pseudo-random sequence of two qubit gates applying it to an arbitrary separable initial state. After sufficient number of two qubit gates we will end up in a random state. Such random protocols have been numerically studied in~\cite{Emerson:03,numer,PRA:07}. Recently~\cite{Plenio}, random protocol for certain gates has been mapped to a Markov chain whose gap determines the convergence rate of a bipartite entanglement. It has been proved that the gap is lower bounded by $\sim 1/n^2$ for CNOT gate and coupling between random qubits, rigorously establishing that only polynomial number of two qubit gates is needed to generate random bipartite entanglement. Numerical investigation~\cite{PRA:07} indicated that the bound is actually $\sim 1/n$, which was furthermore confirmed for random $U(4)$ gate and coupling between random qubits by an analytical study~\cite{Harrow}. 

In the present paper we improve and extend on the bounds in~\cite{Plenio,Harrow} by analytically calculating the exact gap for $U(4)$, CNOT and XY gates and for various coupling topologies like nearest neighbor with open and periodic boundary conditions and for coupling between all qubits. Note that analytical bounds for CNOT and XY gates are not known and that experimentally important case of nearest neighbor coupling has not been treated before. We find that in all cases the gap scales as $\sim 1/n$, meaning that purity reaches level $I \sim \epsilon$ after time scaling as $\sim n \log{(1/\epsilon)}$. Method is easily generalizable to other gates.

As a measure of bipartite entanglement we are going to use purity given by the trace of the square of the reduced density matrix $\rho_A(t)=\trr{B}{\ket{\psi(t)}\bra{\psi(t)}}$. Expanding density matrix $\rho=\ket{\psi(t)}\bra{\psi(t)}$ over products of local Pauli matrices, $\rho=\sum_\mathbf{\alpha}{c_\mathbf{\alpha} \,\, \sigma_1^{\alpha_1} \cdots \sigma_n^{\alpha_n} }$, we get purity $I(t)=\trr{A}{\rho^2_A(t)}=\frac{1}{N^2} \sum_{\mathbf{\alpha} =  \{\alpha_A 0_B \}} c_\alpha^2(t)$. Here $\sigma_i^{\alpha_i}$ denotes Pauli matrix $\alpha_i \in \{0,x,y,z\}$ acting on $i$th qubit, with the convention $\sigma^0=\mathbbm{1}$. We want to calculate time dependence of purity for a protocol consisting of application of a random two-qubit gate $U_{ij}(t)$ acting on $i$th and $j$th qubits at step $t$, $\ket{\psi(t+1)}=U_{ij}(t) \ket{\psi(t)}$. We shall consider two kinds of protocols: (i) $U_{ij}(t)$ is going to be a random $U(4)$ gate, independent for each step, and (ii) $U_{ij}(t)$ is going to be a product of random single qubit gates $V(t)$ and $V'(t)$, independent for each qubit and at each step, and a fixed two qubit gate $W$, $U_{ij}(t)= V_i(t) V'(t)_j W_{ij}$. Averaging over random single qubit unitaries $V$ and $V'$ from $U(2)$ one can arrive at the transformation law of coefficients $c_{\mathbf{\alpha}}$ after one step of the protocol. If a two qubit gate $W$ preserves products of Pauli matrices, {\em i.e.}, if $W$ transforms a product of two Pauli matrices into a product of some other two Pauli matrices (apart from a sign), then the transformation can actually be written for squares of $c_{\mathbf{\alpha}}$ (for details see the original derivation in~\cite{Plenio}), namely,
\begin{equation}
c^2(t+1)=M\, c^2(t),\qquad M=\frac{1}{L}\sum_{ij} M^{(2)}_{ij},
\label{eq:M}
\end{equation}
where by $c^2$ we denote a vector with components $c_{\mathbf{\alpha}}^2$. As one can see, the protocol is described by a Markovian matrix $M$ which is a sum of two-site Markovian matrices $M^{(2)}$ between all coupled qubits. Two qubit matrix $M^{(2)}$ depends only on the two qubit gate $W$ used in the protocol. Its precise form will be given later. Markov matrix $M$ has two eigenvalues equal to $1$, corresponding to invariant states being an identity operator and a uniform mixture of all basis states. We want to calculate the value of the 3rd eigenvalue, $1-\Delta$, where $\Delta$ is the gap of Markovian matrix $M$. If this eigenvalue is nondegenerate purity will asymptotically decay as $I(t) \approx \exp{(-t \Delta)}$ and will reach level $I(\tau)=\epsilon$ after convergence time $\tau=\frac{1}{\Delta}\ln{\frac{1}{\epsilon}}$. In the following we are going to calculate the gap of Markov chain $M$ (\ref{eq:M}) for various 2-qubit gates, thereby obtaining decay time of purity. We will be in particular interested in one-dimensional chains of $n$ qubits with open ($L=n-1$) or periodic ($L=n$) boundary conditions or with coupling between all pairs of qubits. For pedagogical reasons we shall give a detailed derivation for random U(4) gate with other gates being very similar.

{\em U(4) gate--} Let us first have a look at the most symmetric case when two-qubit transformations $U_{ij}(t)$ are independent random U(4) matrices. In this case~\cite{PRA:07,Harrow} two-qubit matrix $M^{(2)}$ is equal to $P_{16 \times 16}$ defined as,
\begin{equation}
P_{m \times m}=\begin{pmatrix}
1 & 0 \\
0 & F
\end{pmatrix},\qquad 
F =\frac{1}{m-1}
\begin{pmatrix}
1 & \ldots & 1\\
\vdots & \ddots & \vdots \\
1 & \ldots & 1
\end{pmatrix}.
\end{equation}
$F$ is here $(m-1) \times (m-1)$ matrix. We defined $P$ to have an arbitrary dimension $m$ because we are actually going to solve a more general situation with a two-site Markov matrix $M^{(2)}$ given by $m \times m$ dimensional matrix $P_{m \times m}$. While dimension $m$ can be arbitrary, we are in particular interested in cases when $P_{m \times m}$ acts on a tensor product of two local bases of dimension $k$, that is $m=k^2$. If we are dealing with qudits with local operator basis of dimension $d^2$ we have $m=d^4$, for instance, $P_{16 \times 16}$ corresponds to qubits and is the one we are most interested in. We are first going to study $P_{16 \times 16}$ and at the end just state results for some other dimensions. 

Any matrix can be decomposed into a sum of tensor products with the minimal number of terms, in quantum information sometimes referred to as the operator Schmidt decomposition~\cite{Nielsen:03},
\begin{equation}
P_{16 \times 16}=\sum_{j=1}^{r} \kappa_j A_j \otimes B_j.
\label{eq:AB}
\end{equation}
Such decomposition can be obtained by singular value decomposition where $\kappa_j$ are positive singular values while $A_j,B_j$ are ``columns'' of unitary transformations occurring in singular value decomposition and act only on a single site. For $P_{16 \times 16}$ such decomposition has only $4$ terms and due to the symmetry $B_j$ can be chosen to be equal to $A_j$. Crucial observation is that all $4 \times 4$ matrices $A_j$ have the same two dimensional kernel (null-space) spanned by vectors $(0,-2/\sqrt{6},1/\sqrt{6},1/\sqrt{6})$ and $(0,0,-1/\sqrt{2},1/\sqrt{2})$. Therefore, using unitary transformation, $A'_j=U A_jU^\dagger$, with
\begin{equation}
U=\frac{1}{2}\begin{pmatrix}
\sqrt{3} & 1/\sqrt{3} & 1/\sqrt{3} & 1/\sqrt{3}\\
-1 & 1 & 1 & 1 \\
0 & -2\sqrt{2/3} & \sqrt{2/3} & \sqrt{2/3}\\
0 & 0 & -\sqrt{2} & \sqrt{2}
\end{pmatrix},
\label{eq:U}
\end{equation}
$A'_j$ have a block form with nonzero elements only in the upper left $2\times 2$ corner. Nontrivial local space is therefore only of dimension $2$! Resumming these $2\times 2$ blocks in $A_j$ and $B_j$ (\ref{eq:AB}) one gets a reduced $P^{\rm red}_{16 \times 16}$ of size $4 \times 4$ which is exactly equal to XY hamiltonian,
\begin{equation}
P^{\rm red}_{16 \times 16}=\frac{1}{2}(\mathbbm{1}\otimes \mathbbm{1}+h_{\rm XY}),\qquad \gamma=\frac{3}{5},h=\frac{4}{5},
\label{eq:P16}
\end{equation}
with
\begin{equation}
h_{\rm XY}=\frac{1+\gamma}{2} \sigma^{\rm x}_i \sigma^{\rm x}_{j}+\frac{1-\gamma}{2} \sigma^{\rm y}_i \sigma^{\rm y}_{j}+h (\frac{1}{2}\sigma^{\rm z}_i+\frac{1}{2}\sigma^{\rm z}_{j}).
\label{eq:hXY}
\end{equation}
Spectrum of Markov chain $M$ (\ref{eq:M}) will therefore be a union of eigenenergies of the full $n$ site XY chain (\ref{eq:P16}) and eigenenergies of all sub-chains obtained by dropping some sites and the corresponding couplings $M^{(2)}_{ij}$ connecting these sites with the rest. These sub-chain eigenenergies come from cases when an operator on the corresponding site comes from the kernel of $A_j$. 

For nearest-neighbor coupling we can use exact results for XY chain in magnetic field~\cite{Katsura:62} to get the largest three eigenvalues of $M$ for $P_{16 \times 16}$. Although solutons are well known, for the sake of completeness and to properly treat boundary conditions we give details in the Appendix. For periodic boundary conditions the resulting gap is
\begin{equation}
\Delta=\frac{2(1-h\cos{(\pi/n)})}{n},
\label{eq:Dpbc}
\end{equation}
with $h=4/5$. For open boundary conditions eigenvalue $1-\Delta$ is doubly degenerate with the gap
\begin{equation}
\Delta=\frac{1-h\cos{(\pi/n)}}{n-1}.
\label{eq:Dobc}
\end{equation}
In both cases the largest three eigenvalues come from the full chain of $n$ sites as the eigenvalues corresponding to sub-chains are strictly smaller. Note that parameters of XY model satisfy $\gamma^2+h^2=1$ for which the ground state becomes doubly degenerate~\cite{Kurmann}, corresponding to two invariant states of our Markov chain. Equations~\ref{eq:Dpbc} and~\ref{eq:Dobc} provide {\em exact expressions} for the gap of Markov chain for periodic or open boundary conditions for any $n$. 

Next, let us proceed with the case when the two-qubit gate is allowed between an arbitrary pair of qubits, that is when the sum in $M$ (\ref{eq:M}) runs over all $L=n(n-1)/2$ distinct pairs. Calculation of the gap, at least in the limit $n \to \infty$, should be easier for such infinite range coupling as one could, for instance, use mean field approximation applicable for infinite dimensional systems. We can actually do better though. First, one notes that for a general XYZ coupling, $h_{\rm XYZ}=J_{\rm x}\sigma^{\rm x}_i \sigma^{\rm x}_{j}+J_{\rm y}\sigma^{\rm y}_i \sigma^{\rm y}_{j}+J_{\rm z} \sigma^{\rm z}_i \sigma^{\rm z}_{j}+\frac{h}{2}(\sigma^{\rm z}_i+\sigma^{\rm z}_{j})$, between all pairs of spins, $M^{\rm red}=\sum_{i<j}(d\, \mathbbm{1}\otimes \mathbbm{1}+h_{\rm XYZ})/L$, we can rewrite this Markov matrix in terms of operators of a total spin, $S_{\alpha}=\frac{1}{2}(\sum_{i=1}^n \sigma^{\alpha}_i)$, $\alpha={\rm x,y,z}$,
\begin{eqnarray}
M^{\rm red}= \frac{2h}{n} S_{\rm z} &+& \frac{4}{n(n-1)} \left( J_{\rm x} S_{\rm x}^2 + J_{\rm y} S_{\rm y}^2+J_{\rm z} S_{\rm z}^2  \right) + \nonumber \\
&+& \left(d-\frac{J_{\rm x}+J_{\rm y}+J_{\rm z}}{n-1} \right)\mathbbm{1},
\label{eq:LMG}
\end{eqnarray}
which is the Lipkin-Meshkov-Glick model (LMG)~\cite{LMG}, solvable by a rather complicated algebraic Bethe ansatz~\cite{Pan}. Note that the dimensionality of the eigenvalue problem has been reduced from $\sim 4^n$ for the original $M$ to $2S+1 \sim n$ for each total spin sector $S$. As we are mainly interested in the scaling of the gap we are going to calculate $\Delta$ in the limit of large $n$. Because largest eigenvalues of $M^{\rm red}$ (\ref{eq:LMG}) come from the sector of maximal spin $S=n/2$ we only have to calculate the difference between the two largest eigenvalues of LMG in this sector. There are various possibilities, perhaps the simplest one being by replacing quantum spin with a classical one, parameterized by a canonical pair $\varphi$ and $\mu=\cos{\theta}$ as $S_{\rm z}=S \mu$, $S_{\rm x}=S\sqrt{1-\mu^2} \cos{\varphi}$ and $S_{\rm y}=S \sqrt{1-\mu^2} \sin{\varphi}$, expanding the resulting classical hamiltonian around its maximum to the lowest order in $\mu$ and $\varphi$, and subsequently quantizing the resulting harmonic oscillator hamiltonian, see {\em e.g.}~\cite{Leyvraz}. We will just state the final result for the gap, which is for our parameters of $U(4)$ gate (\ref{eq:P16}) equal to
\begin{equation}
\Delta=\frac{6}{5n}+{\cal O}(1/n^2).
\label{eq:U4ij}
\end{equation}
Eigenvalue $1-\Delta$ is nondegenerate, while the next one (whose distance to the 2nd one is $\sim 1/n^2$), coming from a sector with $S=n/2-1$, is $n-1$ times degenerate. This degeneracy is simply due to $n-1$ time multiplicity of sector with total spin $S=n/2-1$~\cite{Hamermesh}.

In a similar way one can treat also other dimensions of $P_{m \times m}$. For instance, if local operator basis is of dimension $k=2$, that is for matrix $P_{4 \times 4}$, one can again show equivalence with XY model $h_{\rm XY}$ (\ref{eq:hXY}), this time with parameters $\gamma=1/3$ and $h=2\sqrt{2}/3$. For $P_{9 \times 9}$ we have the equivalence $P^{\rm red}_{9 \times 9} = \frac{1}{2}(\mathbbm{1}\otimes \mathbbm{1}+h_{XY})$ with $\gamma=1/2$ and $h=\sqrt{3}/2$. Physically interesting case is also that of qutrits, for which the equivalence is $P^{\rm red}_{81 \times 81} = \frac{1}{2}(\mathbbm{1}\otimes \mathbbm{1}+h_{XY})$, with parameters $\gamma=4/5$ and $h=3/5$. In all these cases the gap is given by eqs.~(\ref{eq:Dpbc},\ref{eq:Dobc}).

{\em CNOT gate--} If $W$ is CNOT gate two-site transformation matrix $M^{(2)}$ is equal to~\cite{Plenio,PRA:07}
\begin{equation}
M^{(2)}=D\, P_{4 \times 4} \otimes P_{4 \times 4},
\label{eq:Mcnot}
\end{equation}
where $D$ is a permutation matrix giving transformations of products of Pauli matrices by CNOT gate. Its matrix elements are $D_{t'+4c',t+4c}=|\frac{1}{4}\tr{W_{\rm CNOT}\, \sigma^t \sigma^c\, W_{CNOT}^\dagger \,\sigma^{t'} \sigma^{c'}}|=\delta_{t'+4c',f(t+4c)}$ with Kronecker $\delta$, $f(0,1,\ldots,15)=(0, 1, 14, 15, 5, 4, 11, 10, 9, 8, 7, 6, 12, 13, 2, 3)$ and $c$, $t$ denoting target and control qubits, respectively. We proceed along the same lines as for U(4) case. For CNOT we now have only three terms in the sum (\ref{eq:AB}) for $M^{(2)}$. Kernel of matrices $A_j$ and $B_j$ is two dimensional and spanned by the same two vectors as for $P_{16 \times 16}$. Therefore, rotation by unitary $U$, Eq.~(\ref{eq:U}), will bring matrices $A_j$ and $B_j$ to a block form. Because CNOT lacks the full symmetry of U(4) there are now nonzero elements also in the lower left $2 \times 2$ block, in addition to an upper left $2 \times 2$ block. However, as we are interested only in right eigenvectors and eigenvalues we can again take only a sum of products of upper left $2\times 2$ blocks, which is this time equal to the XYZ model,
\begin{equation}
M^{(2)}_{\rm red}=\frac{1}{3}(h_{\rm XY}-\frac{1}{3} \sigma^{\rm z}_i \sigma^{\rm z}_j)+\frac{5}{9}\mathbbm{1}\otimes \mathbbm{1},\quad \gamma=1, h=\frac{4}{3}.
\label{eq:cnot}
\end{equation} 

{\em XY gate--} XY gate is given by $W_{\rm XY}\ket{01}=-\ii \ket{10}, W_{\rm XY}\ket{10}=-\ii \ket{01}, W_{\rm XY}\ket{00}=\ket{00}, W_{\rm XY}\ket{11}=\ket{11}$, and was shown in~\cite{PRA:07} to be faster than CNOT or U(4) for nearest-neighbor couplings. The form of the two-site matrix $M^{(2)}$ is the same as for CNOT (\ref{eq:Mcnot}) with the permutation matrix now given by $D_{t'+4c',t+4c}=\delta_{t'+4c',g(t+4c)}$ with $g(0,1,\ldots,15)=(0, 11, 7, 12, 14, 5, 9, 2, 13, 6, 10, 1, 3, 8, 4, 15)$. Writing $M^{(2)}$ as a tensor sum (\ref{eq:AB}) we have 6 terms with the kernel of each matrix being again at least $2$ dimensional and containing the same two vectors as for $P_{16 \times 16}$. Rotation by $U$ (eq.~\ref{eq:U}) results in matrices of the same structure as for CNOT gate, with the sum of products of nontrivial $2\times 2$ blocks being equal to XYZ model,
\begin{equation}
M^{(2)}_{\rm red}=\frac{2}{3}(h_{\rm XY}+\frac{1}{12} \sigma^{\rm z}_i \sigma^{\rm z}_j)+\frac{7}{18}\mathbbm{1}\otimes \mathbbm{1},\quad \gamma=\frac{1}{2}, h=\frac{2}{3}.
\label{eq:xy}
\end{equation}

For both CNOT and XY gate we obtain equivalence with the XYZ model in magnetic field. For nearest-neighbor coupling the model is not exactly solvable. Nevertheless, due to the existence of two invariant states of $M$ we immediately know that the ground state of a feromagnetic chain is doubly degenerate for these parameters. Chains also have a nonzero energy gap, meaning that the gap $\Delta$ will scale as $\sim 1/n$. For the case of coupling between all pairs of qubits the model is again equivalent to LMG model (\ref{eq:LMG}). Using the same procedure as for $U(4)$ gate we arrive at the gap
\begin{equation}
\Delta=\frac{4}{3n}+{\cal O}(1/n^2),
\label{eq:cnotij}
\end{equation}
which is the same for both CNOT and XY gate~\cite{PRA:07}. Interestingly, the gap is {\em larger} for CNOT and XY gates than for random $U(4)$ gate (\ref{eq:U4ij}), reflecting the fact that many random $U(4)$ gates are only weakly entangling. Regarding degeneracies, for both gates the 2nd and 3rd largest eigenvalues (coming from $S=n/2$) are nondegenerate, while the 4th one (coming from $S=n/2-1$) is $n-1$ times degenerate for the same reason as for $U(4)$. This explains a seeming cutoff-like behavior observed in~\cite{Plenio}. 

{\em Conclusion--}
We have analytically calculated the gap of Markovian chain governing convergence of bipartite entanglement to that of random states. All calculations proceed by mapping transition matrix to various spin models. For $U(4)$ gate and nearest-neighbor coupling we obtained exact expressions for arbitrary $n$. For coupling between all pairs of qubits asymptotically exact expression is obtained for $U(4)$, CNOT and XY gate. In all cases gap scales as $\sim 1/n$. The used method could be also employed for other gates or for qudits.

{\em Appendix--} We want to find eigenenergies for XY model, $H=\sum_i{h^{\rm xy}_{i,i+1}}$ (\ref{eq:hXY}), with periodic or open boundary conditions~\cite{Katsura:62}. Note that for open boundary conditions magnetic field has strength $h/2$ on 1st and last spins. This is important as otherwise a doubly degenerate ground state for $h^2+\gamma^2=1$ splits into an exponentially close doublet. Using standard Wigner-Jordan transformation into fermionic operators $c_j=\sigma_1^{\rm z} \cdots \sigma_{j-1}^{\rm z} (\sigma_j^{\rm x}-\ii \sigma_j^{\rm y})/2$, one gets for periodic boundary conditions in spin variables, $\sigma_{n+1}=\sigma_1$, $H= 2 h\sum_{j=1}^n c_j^\dagger c_j+\sum_{i=1}^{n-1}(c_j c_{j+1}^\dagger +\gamma \, c_j c_{j+1}+{\rm h.c.})-hn +(-1)^{\sum_{k=1}^n 1+c_k^\dagger c_k} \left\{ c_n^\dagger c_1 +\gamma\, c_n^\dagger c_1 + {\rm h.c.} \right\}$. Last term in curly brackets is absent for open boundary conditions. Because parity of the number of fermionic excitations, {\em i.e.} of $\sum_{k=1}^n c_k^\dagger c_k$, is a conserved quantity we can separately diagonalize hamiltonian in even and odd subsectors. In the following we will assume $n$ is even. The only difference between odd and even subsectors is in the sign of the term coupling the last and the 1st fermion. Diagonalization of both can be treated on the same footing by writing $H=h\sum_{j=1}^n (2c_j^\dagger c_j-1)+\sum_{i=1}^{n}(c_j c_{j+1}^\dagger +\gamma \, c_j c_{j+1}+{\rm h.c.})$ and boundary conditions $c_{n+1}=-c_1$ for even subsector and $c_{n+1}=c_1$ for odd subsector. Transforming fermionic operators $c_j$ with a unitary Fourier transformation to reciprocal space fermionic operators $d_k$, $c_j=\frac{{\rm e}^{-\ii \pi/4}}{\sqrt{n}}\sum_{k=1}^n d_k \exp{(\ii \frac{2\pi}{n} k j)}$, we obtain $H=\sum_k h (2 d_k^\dagger d_k-1)+2\cos{(\frac{2\pi}{n}k)d_k d_k^\dagger}+\gamma \sin{(\frac{2\pi}{n}k)}(d_k^\dagger d_{-k}^\dagger + d_{-k} d_k)$. In order to satisfy boundary conditions for fermionic operators $c_j$ the allowed values of $k$ are $k=\frac{1}{2},\frac{3}{2},\ldots,n-\frac{1}{2}$ in even subsector and $k=0,1,\ldots,n-1$ in odd subsector. Finally, Bogoliubov transformation diagonalizes two-dimensional subspace of $d_k$ and $d_{-k}$, resulting in $H=\sum_k \varepsilon_k (2 f_k^\dagger f_k-1)$ with single-particle fermionic excitation energies
\begin{equation}
\varepsilon_k=\sqrt{\left[\cos{(\frac{2\pi}{n}k)}-h \right]^2+\left[ \gamma \sin{(\frac{2\pi}{n}k)} \right]^2}.
\label{eq:ek}
\end{equation}
In the odd subspace two fermionic eigenenergies with no ``$-k$'' partner are $\varepsilon_{k=0}=h-1$ and $\varepsilon_{k=n/2}=h+1$. Ground state of XY model is doubly degenerate~\cite{Kurmann} for $\gamma^2+h^2=1$ and this is the case occurring for Markov chains. Because the spectrum is (for even $n$) symmetric with respect to 0, there are also two maximal eigenstates. These two correspond to two invariant states of $M$ with eigenvalues $\lambda=1$. For $\gamma^2+h^2=1$ eigenmode energies (\ref{eq:ek}) simplify to $\varepsilon_k=1-h\cos{(\frac{2\pi}{n}k)}$. Eigenenergies $E_j$ of XY model are now obtained by filling even/odd number of fermionic modes. There are two eigenenergies $E_{1,2}=n$, coming one from even and one from odd subsector while the second largest eigenenergy is always from even subsector, is nondegenerate, and equal to $E_3=n-4 \varepsilon_{\rm min}=n-4(1-|h|\cos{(\pi/n)})$. The analysis is similar for odd $n$. At the end the same formulas (\ref{eq:ek}) can be used, one only has to take all $\varepsilon_k$ with a negative sign. Largest three eigenenergies are though given by the same $E_{1,2,3}=n,n,n-4(1-|h|\cos{(\pi/n)})$. Finally, for open boundary conditions the three largest eigenenergies are given for an arbitrary $n$ by $E_{1,2,3}=n-1,n-1,n-1-2(1-|h|\cos{(\pi/n)})$, with $E_3$ being doubly degenerate.

\end{document}